\documentstyle [12pt]{article}
\def\ref{\par\noindent\hangindent 20pt}
\def\double{\baselineskip=24pt}

\begin{document}

\titlepage

\double

\begin{center}
{\bf GLOBULAR CLUSTER SYSTEMS AS DISTANCE INDICATORS:}\\
{\bf METALLICITY EFFECTS ON THE LUMINOSITY FUNCTION}
\end{center}

\bigskip

\bigskip

\begin{center}
K{\small EITH} M. A{\small SHMAN} and A{\small LBERTO} C{\small ONTI}\\
Department of Physics and Astronomy,\\
 University of Kansas,\\
 Lawrence,
KS 66045-2151\\
e-mail: ashman@kusmos.phsx.ukans.edu\\
 conti@kusmos.phsx.ukans.edu\\[0.1in]
and\\[0.1in]
S{\small TEPHEN} E. Z{\small EPF}$^1$\\
Department of Astronomy,\\
 University of California,\\
 Berkeley, CA 94720\\
e-mail: zepf@astron.berkeley.edu\\
\end{center}

\bigskip

\bigskip

\begin{center}
To be published in the September 1995 issue of the {\it Astronomical Journal}
\end{center}

\vfill

\noindent
$^1$ Hubble Fellow

\newpage

\begin{center}
{\bf ABSTRACT}
\end{center}

We investigate the universality of the globular cluster luminosity function
(GCLF) and the use of this function as an extragalactic distance indicator.
Previous studies have found
an offset between GCLF distances and
those obtained with other techniques.
We show that this offset can be understood in terms of a metallicity effect.
Specifically, the globular cluster systems used in distance scale studies
have traditionally been those around elliptical galaxies. These systems have
higher mean metallicities than the Milky Way globular cluster system.
Consequently,
the peak of the GCLF in the systems around
ellipticals is significantly fainter in $B$ and $V$ than the GCLF peak
in the Milky Way. We calculate the shift in the peak of the
GCLF relative to the Milky Way globulars
in $B$, $V$, $R$, $I$ and $J$ for a range of globular cluster metallicities.
Applying these corrections, we find good agreement between GCLF
distances and those obtained using the
surface brightness fluctuations method. The similarity between
metallicity-corrected GCLFs suggests that
the underlying mass function of globular cluster systems is remarkably
constant from one galaxy to another.
Our results allow the GCLF to be employed as an improved distance
indicator.

\newpage

\begin{center}
{\bf 1. INTRODUCTION}
\end{center}

The use of globular clusters as extragalactic distance indicators has been
reviewed in detail by Jacoby et al.\ (1992).
The principal technique exploits the roughly Gaussian shape of the
globular cluster luminosity function (GCLF), defined as
the relative number of globular clusters as a function of magnitude
(cf.\ Hanes \& Whittaker 1987; Harris 1988; Harris et al.\ 1991).
Such a distribution can be characterized by just two parameters: the dispersion
$\sigma$ and the peak $M_0$. It is the GCLF peak that is used as a standard
candle.

Clearly, the GCLF technique is only effective if the value of $M_0$
is constant or varies in a predictable manner from one galaxy to another.
The lack of a generally accepted model of globular cluster formation means
that there is no firm physical basis for supposing this to be true, although
there is no shortage of suggestions as to why this might be the case
(e.g.\ Fall \& Rees 1985; Morgan \& Lake 1989;
Murray \& Lin 1989; Ashman 1990; Larson 1990; Ashman \& Zepf
1992; Harris \& Pudritz 1994). The primary evidence that
the GCLF peak is constant is empirical. For instance, the globular cluster
systems of several Virgo ellipticals have GCLF peaks that vary
by less than 0.2 magnitudes (van den Bergh et al.\ 1985; Harris et al.\
1991; Secker \& Harris 1993; Ajhar, Blakeslee \& Tonry 1994).

While current results are generally encouraging, there are two areas of
concern.
The first is the claim that there might be an offset between
the GCLF peaks in the Milky Way and M31 (Secker 1992; Reed et al.\ 1994).
Since these globular cluster systems have the most complete luminosity
functions, a difference between their peaks is worrying. The second potential
problem was summarized neatly by Jacoby et al.\ (1992). These authors noted
that the GCLF technique is more accurate than its error estimates,
but that it yields distances that are offset by 13\% relative to
those obtained using the
surface brightness fluctuations method. The sense of this offset requires that
the GCLF peak in the observed elliptical galaxies is about 0.25 magnitudes
fainter (in $B$) than the peak in the Milky Way (see also Fleming et al.\
1995).

Formation models that attempt to explain the apparent universality of the GCLF
typically predict a mass function or characteristic globular cluster
mass. Given the
uncertainties in the models, it is only sensible to regard the mass and
luminosity functions as having the same form, by tacitly assuming that
the mass-to-light ratio of all globular clusters is the same. However, it
is equally apparent that systematic variations in mass-to-light ratio
 between globular cluster systems will result in different GCLFs, even
if the underlying mass functions are identical. Such variations can be
produced by differences in the age and metallicity of globular cluster
systems.

The main purpose of this paper is to investigate the effects of
variations in the mean metallicity of globular cluster systems on the
GCLF. Jacoby et al.\ (1992) mentioned that this effect might produce
around 0.1 magnitudes of scatter in the GCLF peak from galaxy to galaxy.
However, the galaxies for which GCLF distances have been
obtained are almost exclusively
ellipticals which are known to have globular cluster
systems with higher mean metallicities than the Milky Way and M31 systems
(e.g.\ Harris 1991; Brodie \& Huchra 1991; Zepf, Ashman \& Geisler 1995).
We therefore argue that current distance determinations using the GCLF
suffer from a systematic bias and not simply a scatter resulting
from metallicity-induced variations in the GCLF peak.

Using Worthey's (1994) population synthesis models, we calculate the
shift in the GCLF peak due to metallicity variations. We assume that
the mass function of globular cluster systems is universal and convolve
this mass function with metallicity distributions with a range of
mean metallicities. We find that, for realistic metallicities,
 the shift in the GCLF peak relative to
that of the Milky Way globulars can significantly exceed the 0.1 magnitudes
estimated by Jacoby et al.\ (1992). Most importantly, we find that this
metallicity effect naturally explains the 13\% offset between GCLF
distances and those obtained using the surface brightness fluctuations method.
This suggests that,
in principle, the GCLF may be used successfully as an accurate distance
indicator.

The plan of the paper is as follows. In Section 2, we analyze the GCLFs
of the Milky Way and M31 and find no statistically significant difference
between the peaks of their GCLFs. We also calculate the underlying mass
functions of the two globular cluster systems.
 In Section 3 we calculate the shifts
in the GCLF in $B$, $V$, $R$, $I$ and $J$ for a range of metallicities.
We discuss the effect of this peak shift on distance estimates
in Section 4. We show that accounting for this metallicity effect brings the
GCLF distance scale in line with distances obtained using surface brightness
fluctuations. We also show that our results favor a universal globular cluster
mass function. Finally, in Section 5, we discuss directions
for future work in this area and present our conclusions.

\begin{center}
{\bf 2. THE MILKY WAY AND M31}
\end{center}

Recent work has suggested
the possibility of an offset between the GCLF peak in the Milky
Way and M31 (Secker 1992; Reed et al.\ 1994).
Such an offset would undermine the use of the GCLF peak as a standard candle.
We therefore study the $V$-band GCLFs of these systems in detail to establish
whether a significant difference is present.

The Milky Way data come from the McMaster catalogue (Harris 1994)
and consist of 122 globular clusters with $V$-band luminosities and
spectroscopic metallicities. The metallicities are used in Section 2.2
 below to convert
the luminosity function into a mass function. (Limiting the dataset in this
way excludes twelve clusters with $V$-band luminosities. The inclusion of these
objects slightly
reduces the peak value but the 90\% confidence intervals are virtually
unchanged.)
Data for the M31 globular cluster system are taken from Reed et al.\ (1994).
To convert to absolute magnitudes, we follow these authors in using
an M31 distance modulus of $(m-M)_V = 24.6$, which assumes a reddening
of $E(B-V) = 0.11$. Histograms of the $V$-band GCLFs for the Milky Way and
M31 are presented in Figure 1.

A preliminary statistical analysis of the 122 Milky Way globular clusters
reveals that the five faintest objects are more than 3$\sigma$ from the
mean of the distribution, assuming that the parent distribution is Gaussian.
The traditional technique of employing the peak of
the GCLF as a distance indicator assumes such
a distribution. Since inclusion of the five faint clusters produces a
highly non-Gaussian distribution, they are excluded from the subsequent
analysis. From a practical point of view, these objects are so faint
($M_V \ge -3.30$) that they would not be detected in observations of
extragalactic globular cluster systems. Thus we restrict our attention to
the remaining 117 Milky Way globular clusters.

It is worth noting that the use of the GCLF peak as a standard candle
is not dependent on fitting a particular functional form to the GCLF.
All that is required is a reproduceable method of identifying the peak
(mode) of the GCLF in different datasets. Thus we do not ascribe any
physical importance to the Gaussian form and use it only as a
convenient method of determining a peak value. We return to this
point in Section 2.1.

We analyse the Milky Way and M31
GCLFs using the ROSTAT statistics package (Beers et al.\
1990; Bird \& Beers 1993).
Our results
are summarized in Table 1. The peak value $M_{V0}$ and dispersion $\sigma$
for the two $V$-band GCLFs assume that the parent distributions are Gaussian.
In other words, these quantities are the usual peak and dispersion quoted
in GCLF studies. Table 1 also gives the
robust biweight estimators
of location and scale $C_{BI}$ and $S_{BI}$
 which require no assumption about the parent distribution.
We include the bootstrapped 90\% confidence
limits on both $M_{V0}$ and $C_{BI}$.  In the case of the Milky Way,
the classical estimators give $M_{V0} = -7.33$ and $\sigma = 1.23$
in $V$-band, and
$M_{B0} = -6.50$ and $\sigma = 1.13$ in $B$-band.

Our primary result is that there is no
statistically significant difference between the GCLF peaks in the Milky
Way and M31. This finding is supported by a two-distribution KS test that
indicates the distributions are not significantly different.
Table 1 shows that the 90\%
confidence limits on $M_{V0}$ for the Milky Way and M31 GCLFs overlap. We
find, like previous authors (e.g. Reed et al.\ 1994), that the M31 peak
is brighter, but the difference is not
statistically significant. In fact, if we assume that the peaks of the
parent GCLFs are indistinguishable and that the distributions are Gaussian,
the finite number of datapoints $N$ inevitably produces an uncertainty in
the mean of $\sigma/\sqrt N$. Using the values in Table 1 this corresponds
to 0.11~mag for both the Milky Way and M31. The true uncertainty is
larger than this value because
the two distributions are somewhat non-Gaussian.

Table 1 also
gives the skewness of the two distributions, along with
the corresponding $P$-values.\footnote{
Since the magnitude system assigns increasingly negative numbers to
brighter objects, the skewness of a distribution of magnitudes has the
opposite sign to the same dataset presented in terms of the
logarithm of luminosity. We have therefore quoted the negative of the
value of the skewness
returned by ROSTAT. This ensures consistency between the skewness
quoted for the GCLFs and the corresponding mass functions described in
Section 2.2 below.}
 Both the distributions are skew, rejecting
the hypothesis that the parent distribution is Gaussian. (For M31 the rejection
is only marginally significant.) Interestingly, the distributions are skew
in the opposite sense. This is reflected in the result that the robust
estimator
of location, $C_{BI}$, is brighter than the Gaussian peak $M_{V0}$
in the case of the
Milky Way distribution, but fainter than $M_{V0}$ in M31. One other striking
aspect about the results in Table 1 is that the dispersion (both the Gaussian
and robust measures) is significantly higher for the Milky Way than M31.

\begin{center}
{\it 2.1 Differences in the GCLFs}
\end{center}

While the similarity in $M_{V0}$ of the M31 and Milky Way
distributions is encouraging for
the GCLF distance estimation method, the differences in the shape and
dispersion of the two
GCLFs require further comment.

The dispersions of the GCLFs of the Milky Way and
M31 quoted above and in Table 1 assume that the GCLFs are
complete. There is some question whether this is the case in M31, since there
are no globular clusters in the dataset fainter than $M_V = -5.5$, despite the
presence of such low luminosity globulars in the Milky Way. The addition
of such clusters to the M31 dataset would increase the dispersion of the GCLF
and move the peak to fainter magnitudes. Since we cannot reasonably ``add in''
clusters to the M31 GCLF, we instead truncate the Milky Way GCLF, removing
clusters fainter than $M_V = -5.5$. Applying ROSTAT to this truncated dataset
we obtain $M_{V0} = -7.61$ and $\sigma = 0.95$ (the robust estimators yield
$C_{BI} = -7.57$ and $S_{BI} = 0.97$). The peak of the Gaussian
fit is marginally {\it brighter} than the M31 GCLF and the dispersion is
{\it lower}. This supports the view that the dispersion in the M31 GCLF
has been underestimated due to a non-detection of faint globulars and that
$M_{V0}$ has been pushed to a brighter value by the same effect.

The remaining issue is whether the GCLF is Gaussian.
Recall that we have culled five faint objects from the Milky Way dataset
and that the remaining 117 points constitute a skewed distribution. Thus the
central question is whether it is justifiable to fit a Gaussian to GCLFs
when using them as distance indicators. An important consideration here is
that in galaxies beyond M31, GCLFs are rarely observed much beyond
$M_{V0}$. Thus in practice the best-fitting Gaussian to a GCLF is based
on the brightest 50\% or so of the distribution. Provided the skewness
is being driven by faint clusters, the non-Gaussian nature of the Milky
Way GCLF is not a concern when using the GCLF peak as a standard candle.
We will give a full discussion of this topic and methods of fitting Gaussians
to incomplete GCLFs in a future paper.

As an alternative to the Gaussian, Secker (1992) has suggested the
use of the $t_5$ distribution, which he shows provides a better fit
to several GCLFs.
We feel that an equally valid approach is to use a Gaussian model but
to use a 3$\sigma$ clip to remove outliers. The real problem is that both
the Gaussian and the $t_5$ distributions are symmetric, whereas the
Milky Way GCLF is not. Moreover, for the specific issue of using the
peak of the GCLF as a standard candle, the overall form of the
distribution is unimportant, provided one has a reliable method of
locating the peak.

A different approach has been taken by McLaughlin (1994) who
notes that the Milky Way GCLF is asymmetric and suggests that the {\it mode}
of GCLFs may provide a more reliable standard candle. The difficulty
with this idea is finding a reliable method of locating the mode without
the unattractive step of binning the data. An algorithm for such a
procedure is currently being developed (Ashman, Conti \& Zepf 1995).
These considerations suggest to us that while a Gaussian model is not perfect,
it is currently as good as any alternative, at least until more data demand
a different paramatrization of the GCLF.

\begin{center}
{\it 2.2 Mass Functions}
\end{center}

In order to quantify the effects of metallicity variations on the GCLF, we
need to have a globular cluster mass function. We derive such a mass function
from the 117 globular clusters in the Milky Way dataset described above.
This is achieved by using spectroscopic metallicities for the clusters from
Harris (1994) along with their $V$-band magnitudes.
Using Worthey's (1994) stellar population synthesis models, we use the
metallicities to obtain $V$-band mass-to-light ratios from which we
calculate a mass
for each cluster.

Note that we are primarily interested in {\it differences} in $(M/L)_V$
produced by metallicity variations. We have used Worthey's (1994) models
for 12 Gyr stellar populations, but the {\it form} of the derived mass
function is similar if we use the 17 Gyr models. The masses of
individual clusters are therefore somewhat uncertain (even for the 12 Gyr
models, the $(M/L)_V$ are higher than those inferred from velocity
dispersion measurements), but the relative masses and thus the overall shape
of the mass distribution is more reliable, at least if Milky Way globular
clusters are roughly coeval.

The results of this procedure are summarized in Table 2, where
we consider the distribution of the logarithm of globular cluster masses
to allow a direct comparison to the GCLF. We have also estimated the mass
function of M31 globulars using a similar technique, except that the
metallicities are based on $(B-V)$ colors from Reed et al.\ (1994).
To retain clusters without $B$ magnitudes, we assigned them
the mean $(B-V)$ color of the dataset. This is clearly less than ideal and
will introduce errors into the calculated mass distribution.
Nevertheless, as illustrated by Table 2, the two
mass functions are indistinguishable. As we argued for the GCLF, the peak and
dispersion of the M31 mass distribution is probably biased by the lack of
clusters fainter than $M_V = -5.5$. It is also worth noting that, based on
the skewness, the M31 logarithmic mass distribution is consistent with
Gaussian, while the Milky Way distribution is only marginally inconsistent with
a parent Gaussian distribution.

The similarity of the two mass functions is, in some ways, more important
than the simlarity of the GCLFs. As noted above, the mass distribution
is a more physically meaningful quantity. In Section 4 we give further
arguments in favor of a universal globular cluster mass function.

\newpage

\begin{center}
{\bf 3. SHIFTS IN THE GCLF PEAK DUE TO METALLICITY VARIATIONS}
\end{center}

Jacoby et al.\ (1992) found that the GCLF distance estimation method
is more accurate than its estimated errors, although it gives distances 13\%
larger than the surface brightness fluctuations technique. This result
has the hallmarks of a systematic effect. In this Section, we investigate
the possibility that the higher mean metallicity of the globular cluster
systems around ellipticals, relative to the callibrating Milky Way system,
is responsible for this offset.

Our starting assumption is that the globular cluster mass function is
universal. Using the logarithmic mass function found for the Milky Way
globular cluster system
in the previous section, we simulate GCLFs by convolving this
mass function with a metallicity distribution. That is, each ``cluster''
has a mass and metallicity drawn from an appropriate parent distribution,
so that its luminosity can be calculated from the $(M/L)$-[Fe/H]
relationships given by Worthey (1994).

As in the calculation of the Milky Way globular cluster mass function, we
use the 12 Gyr stellar population models. Note that this procedure generates
GCLFs that are likely to be {\it more} reliable than the mass function.
This is because, in calculating the GCLFs, the physical effect of
importance is the variation in mass-to-light ratio due to metallicity.
We have already noted that Worthey's (1994) mass-to-light ratios are
rather higher than those observed for globular clusters, and that our
choice of using his 12 Gyr models is somewhat arbitrary. However, if we
use the same models to generate the GCLFs, these uncertainties effectively
cancel out, and our results are only dependent on the ability of the
stellar population models to produce accurate {\it relative}
 mass-to-light ratios
for coeval populations of different metallicities.

When the mass and metallicity distributions are convolved,
the resulting
dispersion in the GCLFs is primarily due to the dispersion
in the mass function. A typical dispersion in the metallicity distribution
of globular cluster systems is 0.4 dex or less, roughly corresponding to
a dispersion of 0.15 magnitudes in $V$-band. (The precise value depends
on the absolute metallicity since the $(M/L)_V$-[Fe/H] relation is
non-linear.) The dispersion of the logarithmic mass function is 0.5 dex
(see Table 2), corresponding to 1.2 magnitudes and is therefore responsible for
almost all the dispersion in the GCLF. We checked this explicitly by
simulating $V$-band GCLFs with a single simulated mass function (based
on Milky Way parameters) and four
sets of 100 simulated metallicity distributions, each with
500 datapoints. The simulations are characterized by a mean
metallicity $\mu$ and Gaussian dispersion $\sigma$.
We generated distributions with ($\mu = -1.3$, $\sigma = 0.2$),
($\mu = -1.3$, $\sigma = 0.7$), ($\mu = -0.4$, $\sigma = 0.2$) and
($\mu = -0.4$, $\sigma = 0.7$). We found that while the peak of the GCLF
varied for the different values of $\mu$ as expected, its dispersion
remained constant, even between the cases with $\sigma = 0.2$ and $\sigma =
0.7$.
Moreover, for the sets with the same $\mu$ the GCLF peak was the same for the
two different values of $\sigma$. This is important since it allows us to
use a single characteristic dispersion for the metallicity distribution in our
full simulations, rather than having to worry about a range of dispersions.

Our main simulations also involve 500 datapoints for each case, but now
we randomly draw points both from a logarithmic mass function and a
metallicity distribution. The parent mass function has the Milky Way
parameters given in Table 2. For the metallicity distributions
we again assumed a Gaussian form, but fixed the dispersion at 0.35 dex
for all cases. Based on the results described above, a metallicity
distribution with zero dispersion
would probably generate equally reliable results, but
our approach more closely simulates the observed properties of these
systems. The mean of the metallicity distribution was varied from --1.6 dex to
--0.2 dex in steps of 0.2 dex, with an additional set of simulations
performed for [Fe/H] = --1.35, the mean metallicity of the Milky Way
globular cluster system. This allowed us to calculate GCLFs for a range
of metallicities in five photometric bands: $B$, $V$, $R$, $I$ and $J$.
For each mean metallicity, 100 GCLFs were simulated for each band. Each
resulting GCLF was fit with a Gaussian and the peak and dispersion were
recorded. Thus for each value of metallicity, we obtained 100 values of
the GCLF peak and the corresponding dispersion. The distribution of peak
values was analyzed using ROSTAT to obtain a mean peak value
 and 90\% bootstrapped confidence limits.

In Figures 2 and 3 and Tables 3 and 4 we present our results. Figure 2 shows
the absolute magnitude of the GCLF peak for the range of metallicities
described above, whereas Figure 3 shows the offset in the peak relative to
the GCLF of the Milky Way globular cluster system. (Note that the Milky
Way peak value is obtained from the mean of
our simulations, but the $B$-band and
$V$-band values
are consistent with those determined directly in Section 2.)
The typical error associated with datapoints in both figures
is about 0.11 magnitudes and is dominated by the uncertainty in the peak
of the Milky Way mass function
(or equivalently, the $V$-band GCLF from which it
is derived). The dispersion in the GCLF peaks for each set of parameter
values is small (around 0.02 magnitudes), as is expected for the 500
datapoints of each simulation.
The ROSTAT analysis revealed that the distribution of GCLF peak values for
a given metallicity and band is itself Gaussian-distributed.
Thus for datasets smaller than 100 points, an error of $\sigma/\sqrt N$
should be added in quadrature to the 0.11 magnitudes.

The most striking result apparent from Figures 2 and 3 is that the shift in the
GCLF peak in $B$ and $V$ can be substantial. Two recent studies give median
globular cluster metallicities of --0.56 dex for NGC 3923 (Zepf et al.\
1995) and --0.31 dex for NGC 3311 (Secker et al.\ 1995). The latter
corresponds to a $B$-band shift of about 0.6 magnitudes relative to the
Milky Way. While the globular cluster system of this galaxy is extreme,
characteristic
metallicities for the globular cluster systems of ellipticals are about 0.5 dex
higher than the Milky Way system, corresponding to GCLF peaks that are fainter
by about 0.25 magnitudes in $B$ and 0.15 in $V$. The shifts are sufficient
to have an appreciable effect on distances derived using the GCLF method.

\begin{center}
{\bf 4. IMPLICATIONS FOR GCLF DISTANCES}
\end{center}

The vast majority of galaxies for which GCLF distances have been obtained
are ellipticals. The basic technique involves fitting a Gaussian to the
observed GCLF, determining the apparent magnitude
of the peak, $m_0$, and obtaining a distance modulus by comparing this
peak to the absolute magnitude of the GCLF peak in the Milky Way.
The details of the procedure are described by Jacoby et al.\ (1992).

The results of Section 3 illustrate that previous applications of the GCLF
method are likely to have overestimated galaxy distances, since in most
cases it has been applied to elliptical galaxies with globular cluster systems
with higher mean metallicity than the Milky Way system. The error in the
derived distance modulus due to this effect is simply the shift in the
GCLF peak, $\Delta M_0$, given in Figure 2 and Table 3. Since the distance
is related to the distance modulus by:
$$
d = 10^{(0.2(m-M) + 1)} ~{\rm pc},
\eqno(4.1)
$$
the fractional error in distance produced by an error $\Delta M_0$ in the
distance modulus is:
$$
\frac{ d_1 }{ d_2 } = 10^{0.2 \Delta M_0},
\eqno(4.2)
$$
where $d_2$ is the true distance and $d_1$ is the distance obtained by
assuming the peak of the observed GCLF has the same absolute magnitude as
$M_0$ in the Milky Way. The percentage error in distance can therefore
be written as
$$
\epsilon(d) = 100(10^{0.2 \Delta M_0} - 1)~\%
\eqno(4.3)
$$
where a positive error indicates that the distance has been overestimated.

In Figure 4 we present the percentage error in derived distance as a function
of metallicity for the different photometric bands.
As noted earlier, most of the galaxies with GCLF distances considered by
Jacoby et al.\ (1992) are ellipticals. The GCLFs of these galaxies were
primarily obtained in $B$-band. Assuming a typical mean metallicity of
--0.8 dex for the globular cluster systems of these galaxies
(Harris 1991 and references therein), Table 4
gives $\Delta M_{B0} \approx 0.24$ mag. Substituting this value into equation
(4.3)
we find that this leads to an overestimate in distance of 12\%. This
compares to the 13\% offset found by Jacoby et al.\ (1992) between GCLF
distances and those found using the surface brightness fluctuations technique.
We conclude that all of the offset between these two methods of distance
estimation can be explained by a shift in the GCLF peak due to metallicity.

\newpage

\begin{center}
{\it 4.1 A Case Study: NGC 1399}
\end{center}

A detailed reassessment of GCLF distances that takes into account metallicity
variations is beyond the scope of the present paper. Ideally, one would
correct $M_0$ in each galaxy based on the observed metallicity of its
globular cluster system, but in many cases such metallicities are not
available. However, a study of the shift in $M_0$ in NGC 1399 provides a
useful illustration of the effect.

Bridges et al.\ (1991) studied the GCLF of NGC 1399 in both $B$ and $V$.
They found the peak of the GCLF to occur at $V = 23.75$ and $B = 24.55$
and assumed GCLF peak values in the Milky Way of $M_{V0} = -7.36$ and
$M_{B0} = -6.84$. This leads to distance moduli of $(m-M)_0 = 31.2$ ($V$)
and $(m-M)_0 = 31.5$ ($B$). [Note that these numbers differ slightly from
those quoted by Bridges et al.\ (1991) due to an inconsistency in
the reddening used by these authors. We have used $A_B = 0.0$ throughout
(Burstein \& Heiles 1984) and corrected the above distance moduli accordingly.]

The mean metallicity of the NGC 1399 globular cluster system is
[Fe/H]~$\approx$~--0.75 (e.g.\ Ostrov et al.\ 1993 and references therein).
This leads to a predicted shift in the GCLF peak relative to the Milky Way
of $\Delta M_{V0} \approx 0.18$ and $\Delta M_{B0} \approx 0.26$. Using these
peak
shifts along
with the GCLFs of Bridges et al.\ (1991) and the values for the Milky Way
GCLF peaks found in Section 2, we find distance moduli to NGC 1399 of:
$$
(m-M)_0 = 31.0 \pm 0.35~~(V)
$$
$$
(m-M)_0 = 30.9 \pm 0.40~~(B)
$$
where the errors are based on Bridges et al.\ (1991) estimate of the
uncertainties in the derived GCLF peaks. Note that about half the difference
in the $B$-band distance modulus derived here and by Bridges et al.\ (1991)
arises through our fainter value for $M_{B0}$ in the Milky Way.

Ciardullo et al.\ (1993) find a distance modulus to NGC 1399 of
$30.99 \pm 0.1$ based on the surface brightness fluctuations method.
Clearly our result is in excellent agreement with this value.

\begin{center}
{\it 4.2 A Universal Globular Cluster Mass Function}
\end{center}

Secker \& Harris (1993) assumed a Virgo distance modulus based on the
surface brightness fluctuations scale in order to derive the peak of the
GCLF in four Virgo ellipticals. They concluded that the GCLF peak in these
ellipticals is fainter than the mean of the Milky Way and M31 peak, with
an offset of $\Delta M_{V0} = 0.31 \pm 0.33$. (This is essentially another
way of describing the offset between the GCLF and surface brightness
fluctuations distance scales.) Secker \& Harris (1993) also confirmed
the earlier result of Harris et al.\ (1991) that the scatter in $M_{V0}$
between ellipticals was less than 0.2 magnitudes.

A more recent study by Fleming et al.\ (1995) provides the important
addition of another GCLF of a spiral galaxy (NGC 4565)
to the database. The GCLF of NGC 4565 has a peak comparable to the Milky
Way and M31 in $V$-band, adding weight to the evidence for an offset between
the GCLFs of spirals and ellipticals. Fleming et al.\ (1995) quote GCLF
peaks of $M_{V0} = -7.4 \pm 0.2$ and $M_{V0} = -7.2 \pm 0.1$ for spirals
and ellipticals, respectively, using all current data. Again, our results
indicate that this difference can be accounted for entirely by metallicity
variations.

Since our simulations assume that the globular cluster mass function is
universal,
their success in explaining variations in $M_0$ provide some evidence for
such universality. This is a potentially important result, since it
suggests that globular cluster masses are independent of both metallicity
and environment. Such a finding provides a useful constraint on models of
globular cluster formation. It is possible in principle to confirm the
universality of the globular cluster mass function, either by measuring
globular
cluster metallicities along with GCLFs, or obtaining GCLFs in
metallicity-insensitive bands.

One possible difference between globular cluster mass functions of spirals
and ellipticals is the dispersion. We argued in Section 2 that the lower
dispersion in the GCLF (and hence in the mass function) of M31 relative to the
Milky Way might be a result of incompleteness. However, various studies
have found that the GCLF dispersion for ellipticals is significantly
higher than that of spirals (e.g.\ Harris 1991; Secker \& Harris 1993;
Fleming et al.\ 1995). For a Gaussian fit, the GCLF dispersion is around
1.4 for ellipticals and 1.2 for spirals. Although the dispersion of the
globular cluster metallicity distribution is broader for
ellipticals than spirals, this is unlikely to be responsible for the observed
difference in the GCLFs. As mentioned in Section 2, for realistic
metallicity distributions, the dispersion in the GCLF is almost completely
dominated by the dispersion in the mass function. The GCLF observations
therefore suggest either that
 the globular cluster mass function is intrinsically
broader in ellipticals relative to spirals, or that some other effect is at
work. For instance, Fleming et al.\ (1995) have suggested that in spirals,
the disk produces dynamical evolution of the GCLF, reducing its
dispersion.

        Another possibility is that the globular cluster mass function is
constant, and the GCLF dispersion is inflated by age differences among
the globular clusters around elliptical galaxies.
The bimodal and multimodal color distributions of most elliptical
galaxy globular cluster systems (Zepf \& Ashman 1993; Zepf et al.\ 1995
and references therein) suggest that the globulars formed in two or more
bursts, possibly as a result of galaxy mergers (Ashman \& Zepf 1992).
Irrespective of the mechanism, multimodal color distributions
suggest there are age
differences between the individual populations of globulars.
Although a detailed analysis of all possible age and metallicity
combinations is beyond the scope of this paper, a preliminary analysis
suggests that age is not a promising way to explain the broader dispersion
of elliptical galaxy GCLFs. If the age-metallicity combination is
fixed to reproduce the observed color distribution, the results
tend to be very similar to those found when the color distribution
is assumed to reflect only metallicity differences. This
is a consequence of the similar effect that age and metallicity have
on integrated stellar colors and mass-to-light ratios.

\begin{center}
{\bf 5. CONCLUSIONS}
\end{center}

We have investigated the use of the GCLF as
an extragalactic distance indicator.
The peak of the GCLF can be used as an effective
standard candle for distance measurements.
However, corrections for metallicity differences between globular cluster
systems must be made since objects of the same mass but different
metallicity will have a different luminosity.
Based on a series
of simulations, we have derived metallicity corrections for a range of
photometric bands and metallicities.
Metallicity variations typically produce a shift of about 0.25 magnitudes
in $B$-band between ellipticals and
the Milky Way. This result is supported by observations
of elliptical galaxies which indicate globular clusters of higher
mean metallicity and fainter GCLF peaks relative to the Milky Way.
Our study shows that the offset between the GCLF and surface brightness
fluctuations distance scales can be explained entirely by this metallicity
effect.

Our results strengthen the evidence that the globular cluster mass function
is remarkably similar in different galaxies, provided our interpretation
of the difference in GCLF peaks between galaxies is correct.
This work also suggests that the accuracy of the GCLF peak as
a distance indicator is improved significantly by either obtaining colors,
so that stellar population differences can be explicitly accounted for,
or by observing in bands least affected by metallicity variations,
such as $I$ or $J$.

\bigskip

\bigskip

We are grateful to Terry Bridges, Bill Harris, George Jacoby, Dean
McLaughlin, Jeff Secker and Guy Worthey
 for useful conversations and contributions.
We also thank an anonymous referee for some perceptive and helpful
comments.
K.M.A. acknowledges the financial support of a Fullam/Dudley Award
and a Dunham Grant from the Fund for Astrophysical Research.
A.C. is supported by a bursary from the University of Trieste.
S.E.Z. acknowledges the support of
NASA through grant number
HF-1055.01-93A awarded by the Space Telescope Science Institute, which
is operated by the Association of Universities for Research in
Astronomy, Inc., for NASA under contract NAS5-26555.

\newpage

\begin{center}
{\bf REFERENCES}
\end{center}

\ref{Ajhar, E.A., Blakeslee, J.P., \& Tonry, J.L. 1994, AJ, 108, 2087}

\ref{Ashman, K.M. 1990, MNRAS, 247, 662}

\ref{Ashman, K.M., Conti, A., \& Zepf, S.E. 1995, in preparation}

\ref{Ashman, K.M., \& Zepf, S.E. 1992, ApJ, 384, 50}

\ref{Beers, T.C., Flynn, K., \& Gebhardt, K.  1990, AJ, 100, 32}

\ref{Bird, C.M., \& Beers, T.C. 1993, AJ, 105, 1596}

\ref{Bridges, T.J., Hanes, D.A., \& Harris, W.E. 1991, AJ, 101, 469}

\ref{Brodie, J.P., \& Huchra, J.P. 1991, ApJ, 379, 157}

\ref{Burstein, D., \& Heiles, C. 1984, ApJS, 54, 33

\ref{Ciardullo, R., Jacoby, G.H., \& Tonry, J.L. 1993, ApJ, 419, 479}

\ref{Fall, S.M., \& Rees, M.J. 1985, ApJ, 298, 18}

\ref{Fleming, D.E.B., Harris, W.E., Pritchet, C.J., \& Hanes, D.A.
1995, AJ, 109, 1044}

\ref{Hanes, D.A., \& Whittaker, D.G. 1987, AJ, 94, 906}

\ref{Harris, W.E. 1988, in The Extragalactic Distance Scale, Proceedings
of the ASP 100th Anniversary Symposium, PASPC, Vol. 4, ed. S. van den
Bergh \& C.J. Pritchet (Provo, UT, ASP), p.~231}

\ref{Harris, W.E. 1991, ARAA, 29, 543}

\ref{Harris, W.E. 1994, Catalog of Globular Cluster Parameters, McMaster
University}

\ref{Harris, W.E., Allwright, J.W.B., Pritchet, C.J., \& van den Bergh, S.
1991, ApJS, 76, 115}

\ref{Harris, W.E., \& Pudritz, R.E. 1994, ApJ, 429, 177}

\ref{Jacoby, G.H., Branch, D., Ciardullo, R., Davies, R.L., Harris, W.E.,
Pierce, M.J., Pritchet, C.J., Tonry, J.L., and Welch, D.L. 1992,
PASP, 104, 599}

\ref{Larson, R.B. 1990, PASP, 102, 709}

\ref{McLaughlin, D.E. 1994, PASP, 106, 47}

\ref{Morgan, S., \& Lake G. 1989, ApJ, 339, 171}

\ref{Murray, S.D., \& Lin, D.N.C. 1989, ApJ, 339, 933}

\ref{Ostrov, P., Geisler, D., \& Forte, J.C. 1993, AJ, 105, 1762}

\ref{Reed, L.G., Harris, G.L.H., \& Harris, W.E. 1994, AJ, 107, 555.}

\ref{Secker, J. 1992, AJ, 104, 1472}

\ref{Secker, J., Geisler, D., McLaughlin, D., \& Harris, W.E. 1995,
AJ, 109, 1019}

\ref{Secker, J., \& Harris, W.E. 1993, AJ, 105, 1358}

\ref{van den Bergh, S., Pritchet, C.J., \& Grillmair, C. 1985, AJ, 90, 595}

\ref{Worthey, G. 1994, ApJS, 95, 107}

\ref{Zepf, S.E., \& Ashman, K.M. 1993, MNRAS, 264, 611}

\ref{Zepf, S.E., Ashman, K.M., \& Geisler, D. 1995, ApJ, 443, 570}

\newpage

\begin{center}
{\bf FIGURE CAPTIONS}
\end{center}

\noindent
{\bf Figure 1.} The $V$-band luminosity functions of the globular cluster
systems of the Milky Way (left) and M31 (right). The best-fitting Gaussians
are also shown. (In the case of the Milky Way, the best fit excludes the
5 faint clusters that are removed by a 3$\sigma$ clip.)

\medskip

\noindent
{\bf Figure 2.} The absolute magnitude of the peak of the GCLF in five
photometric bands plotted against mean globular cluster metallicity.
We have used Worthey's (1994) stellar population synthesis models for
a coeval population with an age of 12~Gyr.
 $B$, $V$, $R$, $I$ and $J$ bands are denoted by solid circles, open
circles, open squares, solid triangles and open triangles, respectively.
It is assumed that the underlying mass function is universal
and has the parameters of the Milky Way distribution.

\medskip

\noindent
{\bf Figure 3.} The shift in magnitude $\Delta M_0$
of the GCLF peak plotted against metallicity. Symbols are the same as those in
Figure 2.

\medskip

\noindent
{\bf Figure 4.} The percentage error in estimated
distance resulting from the assumption
that $M_0$ is the same for all GCLFs irrespective of metallicity.
The error is plotted against mean globular cluster metallicity.
Symbols are the same as those in Figure 2.

\end{document}